\documentclass{nature}
\usepackage{epsfig}
\usepackage{amsmath}

\usepackage{amsmath,mathrsfs,amsbsy,color,graphicx,bm,amsthm,amsfonts}
\usepackage{units}
\usepackage{bbm}
\usepackage{times}
\usepackage{dcolumn}
\usepackage{mathrsfs}
\usepackage{amsmath,amssymb,epsfig,float}

\begin{document}
\baselineskip=0.6 cm
\title{Detecting the Curvature of de Sitter Universe with Two Entangled Atoms}
\author{Zehua Tian$^{1\dag}$, Jieci Wang$^{2}$, Jiliang Jing$^{2}$, and Andrzej Dragan$^{1,\star}$}
\maketitle

\begin{abstract}
\baselineskip=0.5 cm

Casimir-Polder interaction arises from the vacuum fluctuations of quantum field that depend on spacetime curvature and thus is spacetime-dependent. Here we show how to use the resonance Casimir-Polder interaction (RCPI) between two entangled atoms to detect spacetime curvature. We find that the RCPI of two static entangled atoms in the de Sitter-invariant vacuum depends on the de Sitter spacetime curvature relevant to the temperature felt by the static observer. It is characterized by a $1/L^2$ power law decay when beyond 
a characteristic length scale associated to the breakdown of a local inertial description of the two-atom system. However, the RCPI of the same setup embedded in a thermal bath in the Minkowski universe is temperature-independent and is always characterized by a $1/L$ power law decay. Therefore, although a single static atom in the de Sitter-invariant vacuum responds as if it were bathed in thermal radiation in a Minkowski universe, using the distinct difference between RCPI of two entangled atoms one can in principle distinguish these two universes.

\end{abstract}

\begin{affiliations}
\item
Institute of Theoretical Physics, University of Warsaw, Pasteura 5, 02-093 Warsaw, Poland

\item
Department of Physics, Key Laboratory of Low Dimensional Quantum Structures and Quantum Control 
of Ministry of Education, and Synergetic Innovation Center for Quantum Effects and Applications, Hunan Normal University, 
Changsha, Hunan 410081, P. R. China
\\
$^\dag$e-mail: zehuatian@126.com
\\
$^\star$e-mail: dragan@fuw.edu.pl

\end{affiliations}

Casimir effect \cite{Casimir, Casimir-Polder} is one of the striking consequences of the fluctuations present in the vacuum state of a quantum field. This effect and related phenomena has attracted great interest in many branches of fundamental physics, including cosmology, statistical mechanics, colloidal physics, as well as material science and nanophysics \cite{Klimchitskaya, Dalvit, Bordag}. Experimental evidence of the Casimir and Casimir-Polder interactions, has been gathered both in the microscopic and macroscopic level \cite{Klimchitskaya, Dalvit} with an unprecedented level of accuracy. This has inpired researchers to study these interactions in more complicated scenarios involving finite temperatures \cite{finite-T1, finite-T2, finite-T3, finite-T4} and configurations out of equilibrium \cite{out-equilibrium1, out-equilibrium2, out-equilibrium3, out-equilibrium4, out-equilibrium5, out-equilibrium6, out-equilibrium7}. The Casimir-Polder interaction has also been used as an effective mean to display the nonlocal properties of field correlations \cite{Passante1, Passante2}, to probe entanglement \cite{Cirone} and detect the Unruh effect \cite{Rizzuto1, Rizzuto2, Rizzuto3}.

It is well known that the relativistic motion of the interacting systems, as well as the curvature of the background spacetime can modify the Casimir-Polder interaction.
Thus, it is in principle imaginable to extract the information about gravity from the Casimir physics. Such a connection between relativistic motion and the Casimir-Polder force between two atoms, as well as the force between an atom and a conducting plate, has been demonstrated in Ref. \cite{Rizzuto1, Rizzuto2, Rizzuto3}. Furthermore, Casimir-Polder-like force of a single two-level atom has been analyzed in Schwarzschild background \cite{Zhang} and de Sitter spacetime \cite{Tian} in order to probe the spacetime curvature. 

De Sitter spacetime is a very simple curved background that enjoys the same degree of symmetry as the Minkowski spacetime, both having ten Killing vectors. More importantly, it is also a good model of our universe in the far past and the far future, as suggested by our current observations and the theory of inflation. It is known that a single particle interacting with a conformally coupled massless scalar field in the de Sitter invariant vacuum state behaves exactly the same way as the one coupled to thermal bath in Minkowski spacetime \cite{Gibbons, Birrell, Polarski, Deser, Zhou-Yu1, Zhou-Yu2, Tian1, Tian2}. It is therefore difficult or impossible to distinguish the de Sitter spacetime from the Minkowski spacetime containing a thermal bath, with the use of a single locally coupled quantum system. In Ref. \cite{Entanglement-de-Sitter} the authors proposed how to use entanglement present in the quantum fields to detect spacetime curvature and showed that using two local particle-detectors interacting with the field can achieve this goal. In the Minkowski spacetime with a quantum field in a thermal state the pair of detectors will be able to extract some entanglement that wouldn't be present in the corresponding scenario involving de Sitter spacetime. Thus, the authors concluded that the two universes can be distinguished by their entangling power. This interesting issue has also been recently reanalyzed in Refs. \cite{Tian, Correlations-de-Sitter, Correlations-de-Sitter1, Other-means, Other-means1}.

In this paper, we propose a new method of spacetime discrimination involving the resonance Casimir-Polder interaction, occuring when one or more atoms are in their excited states and an exchange of real photons between them takes place \cite{resonance-books, resonance-books1}. Our set-up is modeled as two entangled atoms being coupled to a massless scalar field. We compare a scenario, when the field is in the de Sitter-invariant vacuum and is conformally coupled with de Sitter spacetime, with the scenario involving the Minkowski spacetime with a field in a thermal state. Our results show that the resonance interatomic interaction for the de Sitter spacetime case does depend on the spacetime curvature and certain features of RCPI could in principle be used to distinguish de Sitter universe from the thermal Minkowski spacetime.

\begin{methods}

We apply the open quantum system approach introduced by F. Benatti and R. Floreanini~\cite{Benatti} to obtain the effective Hamiltonian of the two atoms, and study the 
interatomic interaction with it. Here let us note that the approach applied in the current paper is different from that in Refs.  \cite{Entanglement-de-Sitter, Benni, Yasusada}, where the window functions are chosen to modulate the interactions between the atoms and field such that the atoms remain causally disconnected, and their evolution can be regarded as unitary except for a finite duration when the interaction is switched on. In our paper, the atoms, interacting with a bath of fluctuating vacuum scalar field, are treated as an open quantum system and that therefore evolve nonunitarily. By tracing over the field degrees of freedom we can derive the master equation that governs the atoms' evolution. Then we are able to examine the dynamics of this open quantum system with the help of the master equation. Besides, our studies are confined in the frame of the two atoms which is regarded as the proper frame, and thus all the physical quantities defined in this frame are spacetime-independent. At this point, let us note that this approach has been extensively used to study the quantum effect \cite{Yu1, Yu2}, such as Hawking effect, and entanglement generation \cite{Hu1} in curved spacetime.

\subsection{Dynamic evolution of two atoms}

Consider two identical and mutually independent atoms that weakly interact with a quantized massless scalar field in its vacuum state. Each of the atoms has two internal energy levels, $\pm\frac{1}{2}\omega_0$, associated with the eigenstates $|e\rangle$ (excited state) and $|g\rangle$ (ground state), respectively. The corresponding Hamiltonian is of the form $H_A^{(\alpha)}=\frac{1}{2}\omega_0\sigma_3^{(\alpha)}$, where the superscript $\alpha$ labels the atom number: $\alpha\in\{1,2\}$, and $\sigma^{(\alpha)}_i$ with $i\in\{1, 2, 3\}$ are Pauli matrices. The total Hamiltonian of the system has the following structure:
\begin{equation}
 H=H_A^{(1)} +H_A^{(2)} +H_F+H_I,
\end{equation}
where $H_F$ represents the free Hamiltonian of the field and the field-atom interaction term, $H_I$, is assumed to be:
\begin{equation}
H_I(\tau)=\mu\left[\sigma^{(1)}_{2} \Phi({\bf x}_1(\tau))+\sigma_{2}^{(2)}\Phi({\bf x}_2(\tau))\right],
\end{equation}
where $\mu$ is the coupling constant that is considered to be small.

Initially, we assume no correlations between the atoms and the external field, therefore the total state of the system is of the form $\rho_{\text{tot}}(0)=\rho(0)\otimes|0\rangle\langle0|$, with $\rho(0)$ being the initial state of the two-atom system, and $|0\rangle$ being the vacuum state of the scalar field. In the frame of atoms, the time evolution of the total system satisfies the von Neumann equation:
\begin{eqnarray}\label{N-equation}
\frac{\partial\rho_{\text{tot}}(\tau)}{\partial\tau}=-i[H(\tau),\rho_{\text{tot}}(\tau)],
\end{eqnarray}
where $\tau$ is the proper time of atoms. We will be interested in the time evolution of the two-atom system, thus by tracing over the field degrees of freedom, i.e., $\rho(\tau)=\text{Tr}_F[\rho_{\text{tot}}(\tau)]$, we can derive the reduced dynamics of the two-atom system. The resulting equation in the weak-coupling limit has the Kossakowski-Lindblad form \cite{Gorini, Lindblad, open}:
\begin{equation}\label{master}
\frac{\partial\rho(\tau)}{\partial\tau}
=-i\big[H_{\rm eff}, \rho(\tau)\big]+{\cal L}[\rho(\tau)],
\end{equation}
with
\begin{equation} \label{effective-Hamiltonian}
H_{\rm eff}=\sum^2_{\alpha=1}H^{(\alpha)}_A-\frac{i}{2}\sum_{\alpha,\beta=1}^2\sum_{i,j=1}^3
H_{ij}^{(\alpha\beta)}\,\sigma_i^{(\alpha)}\,\sigma_j^{(\beta)}\ ,
\end{equation}
and
\begin{equation}
{\cal L}[\rho]
={1\over2} \sum_{\alpha, \beta=1}^2\sum_{i,j=1}^3
 C_{ij}^{(\alpha\beta)}
 \big[2\,\sigma_j^{(\beta)}\rho\,\sigma_i^{(\alpha)}
 -\sigma_i^{(\alpha)}\sigma_j^{(\beta)}\, \rho
 -\rho\,\sigma_i^{(\alpha)}\sigma_j^{(\beta)}\big],
\end{equation}
where $H_{\rm eff}$ is the effective Hamiltonian of the two-atom system. The elements of the matrices $C_{ij}^{(\alpha\beta)}$ and $H_{ij}^{(\alpha\beta)}$ are determined by the Fourier transforms of the field correlation functions:
\begin{equation}\label{green}
\mathrm{}G^{(\alpha\beta)}(\tau-\tau')=\langle\Phi(\tau,\mathbf{x}_{\alpha})\Phi(\tau',\mathbf{x}_\beta)
 \rangle,
\end{equation}
${\cal G}^{(\alpha\beta)}(\lambda)$ and their Hilbert transforms ${\cal K}^{(\alpha\beta)}(\lambda)$, which are respectively given by:
\begin{equation}\label{fourierG}
{\cal G}^{(\alpha\beta)}(\lambda)
=\int_{-\infty}^{\infty} d\Delta\tau\,
 e^{i{\lambda}\Delta\tau}\, G^{(\alpha\beta)}(\Delta\tau)\; ,
\end{equation}
and
\begin{equation}\label{HTR}
{\cal K}^{(\alpha\beta)}(\lambda)
=\frac{P}{\pi i}\int_{-\infty}^{\infty} d\omega\
 \frac{{\cal G}^{(\alpha\beta)}(\omega)}{\omega-\lambda} \;,
\end{equation}
with $P$ being the principal value. Then the elements, $H_{ij}^{(\alpha\beta)}$, in the effective Hamiltonian, $H_{\rm eff}$, can be written explicitly as:
\begin{eqnarray}\label{Hij}
H_{ij}^{(\alpha\beta)}
= A^{(\alpha\beta)}\delta_{ij}-iB^{(\alpha\beta)}\epsilon_{ijk}\,\delta_{3k}-A^{(\alpha\beta)}\delta_{3i}\,\delta_{3j},
\end{eqnarray}
where
\begin{equation}\label{abc1}
\begin{aligned}
A^{(\alpha\beta)}
=\frac{\mu^2}{4}\,[\,{\cal K}^{(\alpha\beta)}(\omega_0)+{\cal K}^{(\alpha\beta)}(-\omega_0)],\\
B^{(\alpha\beta)}
=\frac{\mu^2}{4}\,[\,{\cal K}^{(\alpha\beta)}(\omega_0)-{\cal K}^{(\alpha\beta)}(-\omega_0)].
\end{aligned}
\end{equation}
The elements $C_{ij}^{(\alpha\beta)}$ are given by the equation:
\begin{eqnarray}\label{Cij}
C_{ij}^{(\alpha\beta)}
= \tilde{A}^{(\alpha\beta)}\delta_{ij}-i\tilde{B}^{(\alpha\beta)}\epsilon_{ijk}\,\delta_{3k}-\tilde{A}^{(\alpha\beta)}\delta_{3i}\,\delta_{3j},
\end{eqnarray}
where
\begin{equation}\label{abctilde1}
\begin{aligned}
\tilde{A}^{(\alpha\beta)}
=\frac{\mu^2}{4}\,[\,{\cal G}^{(\alpha\beta)}(\omega_0)+{\cal G}^{(\alpha\beta)}(-\omega_0)],\\
\tilde{B}^{(\alpha\beta)}
=\frac{\mu^2}{4}\,[\,{\cal G}^{(\alpha\beta)}(\omega_0)-{\cal G}^{(\alpha\beta)}(-\omega_0)].
\end{aligned}
\end{equation}

\end{methods}

\section*{Results}

\subsection{Energy-level shifts of two atoms}

Our interest is the effective Hamiltonian of two atoms, $H_{\rm eff}$, from which we can study how the two mutually independent atoms interact with each other through the field medium. Let us note that this Hamiltonian contains two important parts, one is 
$H^{(1)}_A+H^{(2)}_A$, resulting from the internal energy of two isolated atoms, and
another term given by:
\begin{eqnarray}\label{LS}
H_{\rm LS}\equiv-\frac{i}{2}\sum_{\alpha,\beta=1}^2\sum_{i,j=1}^3H_{ij}^{(\alpha\beta)}\sigma_i^{(\alpha)}\,\sigma_j^{(\beta)},
\end{eqnarray}
is analogous to the Lamb shift of the two-atom system resulting from the interaction between the atoms and the external field. In the collective states representation, i.e., the ground state $|G\rangle=|g_1\rangle|g_2\rangle$, the upper state $|E\rangle=|e_1\rangle|e_2\rangle$, the symmetric state $|S\rangle=\frac{1}{\sqrt{2}}(|e_1\rangle|g_2\rangle+|g_1\rangle|e_2\rangle)$, and the antisymmetric state $|A\rangle=\frac{1}{\sqrt{2}}(|e_1\rangle|g_2\rangle-|g_1\rangle|e_2\rangle)$ first introduced by Dicke \cite{Dicke}, the two-atom system behaves as a single four-level system with the above four eigenstates \cite{Ficek}. Thus, by calculating the average values of $H_{\rm LS}$ on the corresponding eigenstates, one can obtain the energy-level shifts of the ground state, the upper state, the symmetric state and the antisymmetric state as:
\begin{eqnarray}\label{energy-level-shifts}
\nonumber
\delta\,E_{G_{LS}}=\langle\,G|H_{\rm LS}|G\rangle&=&-\frac{i}{2}\bigg[H^{12}_{33}+H^{21}_{33}+\sum^3_{i=1}\big(H^{11}_{ii}+H^{22}_{ii}\big)-i\sum^2_{\alpha=1}\big(H^{\alpha\alpha}_{12}-H^{\alpha\alpha}_{21}\big)\bigg],
\\ \nonumber
\delta\,E_{E_{LS}}=\langle\,E|H_{\rm LS}|E\rangle&=&-\frac{i}{2}\bigg[H^{12}_{33}+H^{21}_{33}+\sum^3_{i=1}\big(H^{11}_{ii}+H^{22}_{ii}\big)+i\sum^2_{\alpha=1}\big(H^{\alpha\alpha}_{12}-H^{\alpha\alpha}_{21}\big)\bigg],
\\ \nonumber
\delta\,E_{S_{LS}}=\langle\,S|H_{\rm LS}|S\rangle&=&-\frac{i}{2}\bigg[\sum^3_{i=1}\big(H^{12}_{ii}+H^{21}_{ii}+H^{11}_{ii}+H^{22}_{ii}\big)-2\big(H^{12}_{33}+H^{21}_{33}\big)\bigg],
\\
\delta\,E_{A_{LS}}=\langle\,A|H_{\rm LS}|A\rangle&=&\frac{i}{2}\bigg[\sum^3_{i=1}\big(H^{12}_{ii}+H^{21}_{ii}-H^{11}_{ii}-H^{22}_{ii}\big)\bigg].
\end{eqnarray}

Let us note that expressions \eqref{energy-level-shifts} are quite general and hold for any spacetime backgrounds. The parameters $A$ and $B$ given in Eq.~\eqref{abc1} are relevant to the field correlation functions in Eq.~\eqref{green}, which are along the trajectories of atoms and depend on the spacetime background. Thus, it is expected that the relevant information about the spacetime geometry and motions of atoms is encoded in $A$ and $B$. As a consequence, different types of spacetime could result in different energy-level shifts of the two-atom system. In the following, we will consider that for two static atoms in de Sitter spacetime and in thermal Minkowski spacetime. We are hoping to find 
the difference between these two cases, and thus distinguish these two spacetime with such difference.


\subsection{Resonance Casimir-Polder interaction between two atoms in de Sitter spacetime}\label{section3}

We will be interested in the computation of the field correlation functions of the conformally coupled massless scalar field in de Sitter spacetime. This background is a solution of the Einstein equations with the cosmological constant $\Lambda$, and it can be conveniently represented as the surface of the hyperboloid:
\begin{eqnarray}
z_0^2-z_1^2-z_2^2-z_3^2-z_4^2=-\alpha^2,
\end{eqnarray}
embedded in the five dimensional Minkowski spacetime with the metric \cite{Birrell}:
\begin{eqnarray}
ds^2=dz_0^2-dz_1^2-dz_2^2-dz_3^2-dz_4^2,
\end{eqnarray}
where $\alpha=\sqrt{3/\Lambda}$. By applying the following parametrization:
\begin{eqnarray}
\begin{aligned}
&z_0=\sqrt{\alpha^2-r^2}\sinh{t/\alpha},\\
&z_1=\sqrt{\alpha^2-r^2}\cosh{t/\alpha},\\
&z_2=r\cos\theta,\\
&z_3=r\sin\theta\cos\phi,\\
&z_4=r\sin\theta\sin\phi,
\end{aligned}
\end{eqnarray}
we can obtain the static de Sitter metric:
\begin{eqnarray}
\nonumber
\text{d}s^2=\biggl(1-\frac{r^2}{\alpha^2}\biggr)\text{d}{t}^2-\biggl(1-\frac{r^2}{\alpha^2}\biggr)^{-1}\text{d}r^2-r^2(\text{d}\theta^2+\sin^2\theta \text{d}\phi^2).
\\
\end{eqnarray}
Obviously, there is a coordinate singularity at $r=\alpha$ where the so called cosmological horizon is. Note that in curved spacetime, a delicate issue arises of how to determine the vacuum state of the quantum field. Here we 
choose the de Sitter-invariant vacuum state as the state of the conformally coupled massless scalar field, since it is an analogous state to the Minkowski vacuum 
in flat spacetime, and it is considered to be a natural vacuum \cite{Allen}. The corresponding Wightman function takes the form~\cite{Birrell, Polarski}:
\begin{eqnarray}\label{green1}
G^+(x,x')=-\frac{1}{4\pi^2} \frac{1}{(z_0-z_0')^2-\Delta\,z^2-i \epsilon},
\end{eqnarray}
where $\Delta\,z^2=(z_1-z_1')^2+(z_2-z_2')^2+(z_3-z_3')^2+(z_4-z_4')^2$ and $\epsilon$ is an infinitesimal constant. 
We assume that the two static atoms we considered are held at the positions $(r, \theta, \phi)$ and $(r, \theta', \phi)$, respectively. To calculate the corresponding Wightman functions for these two spacetime points, we submit the trajectories of the atoms into Eqs. (\ref{green}) and (\ref{green1}), then we obtain:
\begin{eqnarray}\label{Gaa}
\nonumber
G^{(11)}(x, x^\prime)=G^{(22)}(x,x')&=&-\frac{1}{4\pi^2}\bigg[(\sqrt{\alpha^2-r^2}\sinh\,t/\alpha-\sqrt{\alpha^2-r^2}\sinh\,t^\prime/\alpha)^2
\\  \nonumber
&&-(\sqrt{\alpha^2-r^2}\cosh\,t/\alpha-\sqrt{\alpha^2-r^2}\cosh\,t^\prime/\alpha)^2-i\epsilon\bigg]^{-1}
\\ \nonumber
&=&-\frac{1}{4\pi^2}\frac{1}{2(\sqrt{\alpha^2-r^2})^2\cosh\big[\frac{t-t^\prime}{\alpha}\big]-2(\sqrt{\alpha^2-r^2})^2-i\epsilon}
\\ \nonumber
&=&-\frac{1}{16\pi^2(\sqrt{\alpha^2-r^2})^2\sinh^2\big(\frac{t-t^\prime}{2\alpha}-i\epsilon\big)}
\\
&=&-\frac{1}{16\pi^2\kappa^2\sinh^2(\frac{\Delta\tau}{2\kappa}-i\epsilon)},
\end{eqnarray}
and
\begin{eqnarray}\label{Gab}
\nonumber
G^{(12)}(x, x^\prime)&=&G^{(21)}(x, x^\prime)
\\   \nonumber
&=&-\frac{1}{4\pi^2}\bigg[(\sqrt{\alpha^2-r^2}\sinh\,t/\alpha-\sqrt{\alpha^2-r^2}\sinh\,t^\prime/\alpha)^2
\\  \nonumber
&&-(\sqrt{\alpha^2-r^2}\cosh\,t/\alpha-\sqrt{\alpha^2-r^2}\cosh\,t^\prime/\alpha)^2-(r\cos\theta-r\cos\theta^\prime)^2
\\  \nonumber
&&-(r\sin\theta\cos\phi-r\sin\theta^\prime\cos\phi)^2-(r\sin\theta\sin\phi-r\sin\theta^\prime\sin\phi)^2-i\epsilon\bigg]^{-1}
\\ \nonumber
&=&-\frac{1}{4\pi^2}\frac{1}{2(\sqrt{\alpha^2-r^2})^2\cosh\big[\frac{t-t^\prime}{\alpha}\big]-2(\sqrt{\alpha^2-r^2})^2+2r^2(\cos(\theta-\theta^\prime)-1)-i\epsilon}
\\ \nonumber
&=&-\frac{1}{16\pi^2\bigg[(\sqrt{\alpha^2-r^2})^2\sinh^2\big(\frac{t-t^\prime}{2\alpha}-i\epsilon\big)-r^2\sin^2(\frac{\theta-\theta^\prime}{2})\bigg]}
\\
&=&-\frac{1}{16\pi^2\kappa^2}\frac{1}{\sinh^2(\frac{\Delta\tau}{2\kappa}-i\epsilon)-\frac{r^2}{\kappa^2}\sin^2\frac{\Delta\theta}{2}}.
\end{eqnarray}
where we have used the definitions: $\kappa=\sqrt{g_{00}}\alpha=\sqrt{1-r^2/\alpha^2}\alpha=\sqrt{\alpha^2-r^2}$, and $\Delta\tau=\tau-\tau^\prime=\sqrt{g_{00}}\Delta\,t=\sqrt{g_{00}}(t-t^\prime)$ with $\tau$ being the proper time of the atom. Then, through the contour integral we can calculate the Fourier transforms of the field correlation functions shown in Eqs. \eqref{Gaa} and \eqref{Gab}, which
are given by:
\begin{eqnarray}\label{CGaa}
\nonumber
{\cal\,G}^{(11)}(\lambda)={\cal\,G}^{(22)}(\lambda)&=&\int^\infty_{-\infty}\frac{-1}{16\pi^2\kappa^2\sinh^2(\frac{\Delta\tau}{2\kappa}-i\epsilon)}e^{i\lambda\Delta\tau} d\Delta\tau
\\ \nonumber
&=&2\pi\,i\times\sum^{\infty}_{n=0}\frac{\lambda}{4\pi^2i}e^{-2n\pi\kappa\lambda}
\\ 
&=&\frac{1}{2\pi}\frac{\lambda}{1-e^{-2\pi\kappa\lambda}},
\end{eqnarray}
and 
\begin{eqnarray}\label{CGab}
\nonumber
{\cal\,G}^{(12)}(\lambda)&=&{\cal\,G}^{(21)}(\lambda)
\\ \nonumber
&=&\int^\infty_{-\infty}\frac{-1}{16\pi^2\kappa^2}\frac{1}{\sinh^2(\frac{\Delta\tau}{2\kappa}-i\epsilon)-\frac{r^2}{\kappa^2}\sin^2\frac{\Delta\theta}{2}} d\Delta\tau
\\ \nonumber
&=&2\pi\,i\times\sum^{\infty}_{n=0}
\frac{e^{-2n\pi\kappa\lambda}}{16\bigg(\pi^2r\sqrt{1+\frac{r^2\sin^2\big(\frac{\Delta\theta}{2}\big)}{\kappa^2}}\bigg)\sin\big(\frac{\Delta\theta}{2}\big)}
\\  \nonumber
&&\times\bigg\{\exp\bigg[-2i\kappa\lambda\sinh^{-1}\bigg[\frac{r\sin\big(\frac{\Delta\theta}{2}\big)}{\kappa}\bigg]\bigg]
-\exp\bigg[2i\kappa\lambda\sinh^{-1}\bigg[\frac{r\sin\big(\frac{\Delta\theta}{2}\big)}{\kappa}\bigg]\bigg]\bigg\}
\\
&=&\frac{1}{2\pi}\frac{\lambda}{1-e^{-2\pi\kappa\lambda}}f(\lambda,L/2),
\end{eqnarray}
where $n\in\{\mathbf{Z}\}$, $f(\lambda, z)=\frac{\sin\big[2\kappa\lambda\sinh^{-1}(z/k)\big]}{2z\lambda\sqrt{1+z^2/\kappa^2}}$, and $L=2r\sin(\Delta\theta/2)$ is the usual Euclidean distance between the two points $(r,~\theta,~\phi)$ and $(r,~\theta',~\phi)$, i.e., the distance between the two static atoms in de Sitter spacetime. Consequently, using the results in Eqs. \eqref{CGaa} and \eqref{CGab} together with Eq. \eqref{HTR}, it is found that the Hilbert transforms are given by:
\begin{eqnarray}\label{Hilbert-transforms}
\nonumber
{\cal K}^{(11)}(\omega_0)={\cal K}^{(22)}(\omega_0)
&=&\frac{1}{2\pi^2i}P\int^\infty_{-\infty}d\omega\frac{1}{\omega-\omega_0}\frac{\omega}{1-e^{-2\pi\kappa\omega}},
\\ \nonumber
{\cal K}^{(12)}(\omega_0)={\cal K}^{(21)}(\omega_0)
&=&\frac{1}{2\pi^2i}P\int^\infty_{-\infty}d\omega\frac{1}{\omega-\omega_0}\frac{\omega}{1-e^{-2\pi\kappa\omega}}
\\
&&\times\,f(\omega,L/2).
\end{eqnarray}
Plugging the Hilbert transforms into Eqs.~(\ref{Hij}) and (\ref{abc1}), we obtain:
\begin{eqnarray}\label{HAB}
\nonumber
&&H_{ij}^{(11)}=H_{ij}^{(22)}
=A_1\,\delta_{ij}-iB_1\epsilon_{ijk}\,\delta_{3k}-A_1\delta_{3i}\,\delta_{3j},
 \\
&&H_{ij}^{(12)}=H_{ij}^{(21)}
=A_2\,\delta_{ij}-iB_2\epsilon_{ijk}\,\delta_{3k}-A_2\delta_{3i}\,\delta_{3j},
\end{eqnarray}
where
\begin{eqnarray}\label{ABC}
\nonumber
A_1&=&\frac{\mu^2P}{8\pi^2i}\int^\infty_{-\infty}d\omega\bigg(\frac{\omega}{\omega-\omega_0}+\frac{\omega}{\omega+\omega_0}\bigg)\frac{1}{1-e^{-2\pi\kappa\omega}},
\\  \nonumber
B_1&=&\frac{\mu^2P}{8\pi^2i}\int^\infty_{-\infty}d\omega\bigg(\frac{\omega}{\omega-\omega_0}-\frac{\omega}{\omega+\omega_0}\bigg)\frac{1}{1-e^{-2\pi\kappa\omega}},
\\  \nonumber
A_2&=&\frac{\mu^2P}{8\pi^2i}\int^\infty_{-\infty}d\omega\bigg(\frac{\omega}{\omega-\omega_0}+\frac{\omega}{\omega+\omega_0}\bigg)\frac{1}{1-e^{-2\pi\kappa\omega}}f(\omega,L/2),
\\  \nonumber
B_2&=&\frac{\mu^2P}{8\pi^2i}\int^\infty_{-\infty}d\omega\bigg(\frac{\omega}{\omega-\omega_0}-\frac{\omega}{\omega+\omega_0}\bigg)\frac{1}{1-e^{-2\pi\kappa\omega}}f(\omega,L/2).
\\
\label{abc2}
\end{eqnarray}

Let us now proceed with the study of the Casimir-Polder potential between the two atoms. According to Eqs.~(\ref{HAB}) and (\ref{ABC}), the terms $H^{\alpha\beta}_{ij}$ with $\alpha=\beta$ have no contribution to the interatomic interaction energy, since such terms are independent of the distance, $L$, between the two atoms. As a consequence, 
from the energy shifts of the ground state and the upper state cases given by Eq.~\eqref{energy-level-shifts}, we can see that there is no interatomic interaction between the uncorrelated two atoms in the second-order perturbation theory. However, we find that for both symmetric and antisymmetric entangled states cases, there are terms $H^{\alpha\beta}_{ij}$ with $\alpha\ne\beta$ in the energy shifts of the two atoms, meaning that the interatomic interactions definitely exist in such cases. For these two entangled state cases, the corresponding energy shifts are given by:
\begin{eqnarray}\label{SALS}
\nonumber
\delta\,E_{S_{LS}}&=&-\frac{\mu^2}{4\pi^2}\int^\infty_0d\omega\bigg(\frac{\omega}{\omega-\omega_0}+\frac{\omega}{\omega+\omega_0}\bigg)\big[f(\omega,L/2)+1\big],
\\
\delta\,E_{A_{LS}}&=&\frac{\mu^2}{4\pi^2}\int^\infty_0d\omega\bigg(\frac{\omega}{\omega-\omega_0}+\frac{\omega}{\omega+\omega_0}\bigg)\big[f(\omega,L/2)-1\big].
\end{eqnarray}
It is obvious that in Eq. \eqref{SALS} the term $\int^\infty_0d\omega\big(\frac{\omega}{\omega-\omega_0}+\frac{\omega}{\omega+\omega_0}\big)$ is divergent. However,  this divergence can be removed by taking a cutoff on the upper limit of the integral using Bethe's Method~\cite{Bethe1, Bethe2}. At this point, let us note that the similar processes have been investigated in Refs. \cite{Zhou1, Zhou2}, where energy shift of a two-level atom has been studied in curved spacetime with the formalism developed by Dalibard, Dupont-Roc, and Cohen-Tannoudji \cite{DDC, DDC1}. Besides, it is needed to point out that this integral term also contains no $L$ and thus it is insignificant when we take the derivative
of it with respect to $L$ to calculate the Casimir-Polder force between the two atoms. Due to that, we can rewrite the interatomic interaction for the symmetric and antisymmetric entangled states cases as:
\begin{eqnarray}
\nonumber
\delta\,E_S&=&-\frac{\mu^2}{4\pi^2}\int^\infty_0d\omega\bigg(\frac{\omega}{\omega-\omega_0}+\frac{\omega}{\omega+\omega_0}\bigg)f(\omega,L/2),
\\
\delta\,E_A&=&\frac{\mu^2}{4\pi^2}\int^\infty_0d\omega\bigg(\frac{\omega}{\omega-\omega_0}+\frac{\omega}{\omega+\omega_0}\bigg)f(\omega,L/2).
\end{eqnarray}
The integral in the equations above can be evaluated analytically, resulting in the following expressions:
\begin{eqnarray}\label{ESR}
\nonumber
\delta\,E_S&=&-\frac{\mu^2}{4\pi}\frac{1}{L\sqrt{1+(L/2\kappa)^2}}\cos\bigg(2\omega_0\kappa\sinh^{-1}\big(\frac{L}{2\kappa}\big)\bigg),
\\
\delta\,E_A&=&\frac{\mu^2}{4\pi}\frac{1}{L\sqrt{1+(L/2\kappa)^2}}\cos\bigg(2\omega_0\kappa\sinh^{-1}\big(\frac{L}{2\kappa}\big)\bigg).
\end{eqnarray}
It can be seen that the results depend on the choice of the background metric through the parameter $\kappa=\sqrt{g_{00}}\alpha$. Therefore, the parameters of the de Sitter spacetime can in principle be probed using a pair of atoms interacting via the resonance Casimir-Polder interaction. It is interesting that the response of the single detector \cite{Gibbons, Birrell, Deser} in terms of the spontaneous emission rate, energy-level shift, and geometric phase \cite{Zhou-Yu1, Zhou-Yu2, Tian1, Tian2} in de Sitter spacetime, shows that the detector 
seems as if it were immersed in a thermal bath with the temperature $T=1/2\pi\kappa$. However, the resonance interatomic interactions here manifest non-thermally, carrying no signatures of thermal fluctuations.

In order to investigate the detailed behavior of the RCPI in de Sitter spacetime, let us notice that a characteristic length scale in our problem is $\kappa$. For distances smaller than $\kappa$, it is possible to find a local inertial frame where all the laws of physics are the same with that in Minkowski spacetime. On the other hand, when the considered distances are larger than $\kappa$, the curvature of de Sitter spacetime may play a nontrivial role. For that reason we will focus on the RCPI for distances $L$ between the detectors large enough for the spacetime curvature to have an effect. Alternatively we will also consider the RCPI for very small $L$, when the effect of spacetime curvature can be neglected and the results should be essentially the same, as obtained in Minkowski spacetime.

In the limit of $L\gg\kappa$, i.e., when the two-atom system is near the cosmological horizon, the RCPI given by Eq.~\eqref{ESR} can be written as:        
\begin{eqnarray}\label{ESgg}
\nonumber
\delta\,E_S&=&-\frac{\mu^2}{2\pi}\frac{\kappa}{L^2}\cos\bigg(2\omega_0\kappa\log\bigg(\frac{L}{\kappa}\bigg)\bigg),
\\
\delta\,E_A&=&\frac{\mu^2}{2\pi}\frac{\kappa}{L^2}\cos\bigg(2\omega_0\kappa\log\bigg(\frac{L}{\kappa}\bigg)\bigg),
\end{eqnarray}
and in the limit $L\ll\kappa$ we have: 
\begin{eqnarray}\label{ESll}
\nonumber
\delta\,E_S&=&-\frac{\mu^2}{4\pi}\frac{1}{L}\cos(\omega_0L),
\\
\delta\,E_A&=&\frac{\mu^2}{4\pi}\frac{1}{L}\cos(\omega_0L).
\end{eqnarray}
We can see that in the flat spacetime scenario given by Eq.~\eqref{ESll}, the correction to the energy varies with the interatomic distance as  $L^{-1}$, while in the de Sitter case, given by Eq.~\eqref{ESgg}, the energy decreases as $L^{-2}$. This shows that the resonance interatomic interactions bear a signature of spacetime curvature. We also point out that the pre-factor in Eq.~\eqref{ESgg} explicitly depends on the parameter $\kappa$ associated with the temperature $T=1/2\pi\kappa$ that is felt by static observers in de Sitter spacetime. Let us note that the temperature $T=1/2\pi\kappa$ actually can be written as $T=\sqrt{T^2_f+T^2_a}$. Here $T_f=\frac{1}{2\pi\alpha}$ is the Gibbons-Hawking temperature, and $T_a=a/2\pi$ is the Unruh temperature with $a=\frac{r}{\alpha^2}(1-\frac{r^2}{\alpha^2})^{-1/2}$ being the proper acceleration of static atom \cite{Gibbons, Birrell, Deser, Zhou-Yu1, Zhou-Yu2, Tian1, Tian2}.  Let us note that both $T_f$ and $T_a$ are associated with the curvature of de Sitter spacetime, i.e., $R=12/\alpha^2$ \cite{Birrell}. If the curvature $R$ were zero, i.e., $\alpha\rightarrow\infty$, both $T_f$ and $T_a$ vanish and then the RCPI is reduced to the inertial case shown in Eq. \eqref{ESll}. However, when $a=0$, i.e., the atoms are located at $r=0$, the ``kinematics" of the atoms has no contribution to the RCPI, but it is still related to the spacetime curvature due to the Gibbons-Hawking effect.
Thus, in this regard, Eq.~\eqref{ESgg} implies that it is possible to single out metric effects associated to the curvature of de Sitter spacetime.

In order to compare the results given above with that corresponding to the thermal Minkowski spacetime scenario, we consider the RCPI between two static entangled atoms in Minkowski spacetime, coupled to a massless scalar field in a thermal state characterized by the temperature $T=1/2\pi\kappa$. For this case, the field correlation functions are given by:
\begin{eqnarray}
G^{(11)}(x,x')=G^{(22)}(x,x')
=-\frac{1}{4\pi^2}\sum^{+\infty}_{n=-\infty}\frac{1}{(\Delta\tau-in/T-i\epsilon)^2},
\end{eqnarray}
and
\begin{eqnarray}
\nonumber
G^{(12)}(x,x')&=&G^{(21)}(x,x')
\\ 
&=&-\frac{1}{4\pi^2}\sum^{+\infty}_{n=-\infty}\frac{1}{(\Delta\tau-in/T-i\epsilon)^2-L^2},
\end{eqnarray}
where $\Delta\tau=t-t^\prime$ with $t$ being the proper time of the static atoms in flat spacetime, and $L=2r\sin(\Delta\theta/2)$ denotes the distance between the two atoms. From these correlation functions we can carry out an analogous computation of the RCPI between the two static atoms in the thermal Minkowski spacetime, obtaining:
\begin{eqnarray}\label{ETM}
\nonumber
\delta\,E_{S_M}&=&-\frac{\mu^2}{4\pi}\frac{1}{L}\cos(\omega_0L),
\\
\delta\,E_{A_M}&=&\frac{\mu^2}{4\pi}\frac{1}{L}\cos(\omega_0L).
\end{eqnarray}
Interestingly, these interatomic interactions do not depend on the temperature of the thermal bath, and they are identical to that of two inertial atoms shown in Eq. (\ref{ESll}). We also stress that these interatomic interactions are quite different from the results in Eq. (\ref{ESR}), which means the RCPI for the de Sitter spacetime case and that for the thermal Minkowski spacetime case behave differently. In particular, when the distance between two atoms $L\gg\,\kappa$, the curvature of de Sitter spacetime will strongly affect the nature 
of the field correlation functions $\mathrm{}G^{(\alpha\beta)}(\tau-\tau')$, ultimately leading to the novel power law behavior, i.e., $\sim1/L^2$, of the RCPI between two atoms. However, the RCPI for the thermal Minkowski case behaves with power law $1/L$. Because of the difference of the RCPI, correspondingly, the resonance Casimir-Polder force between the atoms should behave quite differently with the change of distance $L$. Such force in de Sitter spacetime will decrease more quickly than that for the thermal Minkowski spacetime case as $L$ increases. This quite different power law could be used as a criterion to determine the nature of these two universes. Therefore, two entangled atoms in principle can be used to discriminate between two alternative universes, generally speaking, indistinguishable with just a single atom: a thermal Minkowski spacetime or de Sitter spacetime.

\section*{Conclusions and Discussions} 

We used the open quantum system approach to derive the dynamics of the two-atom system, in particular, its effective Hamiltonian. This allows us to compute the RCPI between two entangled atoms. We calculated such RCPI in de Sitter-invariant vacuum and that in flat spacetime with field in the thermal state. We find that the former depends on the de Sitter spacetime curvature relevant to the temperature felt by the static observer and is characterized by a $1/L^2$ power law decay when beyond  a characteristic length scale associated to the breakdown of a local inertial description of the two-atom system. However, the latter is temperature-independent and is always characterized by a $1/L$ power law decay.
Therefore, although de Sitter spacetime and the thermal Minkowski spacetime share a lot of the same properties and can not be distinguished by a single probe, by examining the RCPI between two entangled atoms it is in principle possible to discriminate these two spacetimes.

A similar task can be accomplished by examining the generation of entanglement \cite{Entanglement-de-Sitter} between two initially uncorrelated static atoms. In such a scenario, the two detectors are required to be placed beyond each other's cosmic horizons (in the de Sitter case) therefore the entanglement that is possible to extract is extraordinarily small \cite{Entanglement-de-Sitter}. On the other hand, our proposal does not involve vacuum entanglement extraction and uses feasible amounts of inter-atomic entanglement. Moreover, the requirement for the location of two atoms is much weaker. Our results showed that if spacetime is curved, i.e., in de Sitter universe, the RCPI is characterized by a $1/L^2$ power law decay when $L\gg\kappa$, while this interaction is always proportional to $1/L$ in flat spacetime, no matter whether the field state is thermal or not. In this regard, the criterion proposed in this work seems to be more practical.

\begin{addendum}

\item [Acknowledgement]

Z. Tian and A. Dragan thank for the financial support to the National Science Center, Sonata BIS Grant No. DEC-2012/07/E/ST2/01402. 
J. Jing and J. Wang are supported by the National Natural Science Foundation of China under Grant No. 11475061 and No. 11675052; 
and Construct Program of the National Key Discipline.

\item [Author Contributions]
Z. T. made the main calculations.  J. W., J. J., and A. D. discussed the results, Z. T. wrote the paper with assistances from A. D. and other authors.

\item [Competing Interests]
The authors declare that they have no competing financial interests.

\item [Correspondence]
Correspondence and requests for materials should be addressed to
Z. T., and A. D.
\end{addendum}


\parskip=20 pt

\begin{thebibliography}{99}
\bibitem{Casimir}
Casimir, H. B. G. On the attraction between two perfectly conducting plates. \emph{Proc. K. Ned. Akad. Wet.} {\bf 51}, 793 (1948).

\bibitem{Casimir-Polder}
Casimir, H. B. G. \& Polder, D. The Influence of Retardation on the London-van der Waals Forces. \emph{Phys. Rev.} {\bf 73}, 360 (1948).

\bibitem{Klimchitskaya}
Klimchitskaya, G. L., Mohideen, U. \& Mostepanenko, V. M.  
The Casimir force between real materials: Experiment and theory. \emph{Rev. Mod. Phys.} {\bf 81}, 1827 (2009).

\bibitem{Dalvit}
Dalvit, D.,  Milonni, P.,  Roberts, D. \& Rosa, F.  \textit{Casimir Physics.} Lecture Notes in Physics Vol. 834, (Springer, New York, 2011).


\bibitem{Bordag}
Bordag, M., Klimchitskaya, G. L.,  Mohideen, U. \& Mostepanenko, V. M.  \textit{Advances in the Casimir Effect}, (Oxford University Press, New York, 2009).

\bibitem{finite-T1}

Barton, G. Long-range Casimir-Polder-Feinberg-Sucher intermolecular potential at nonzero temperature. \emph{Phys. Rev. A} {\bf 64}, 032102 (2001).
\bibitem{finite-T2}

Passante, R. \& Spagnolo, S. Casimir-Polder interatomic potential between two atoms at finite temperature and in the presence of boundary conditions. 
\emph{Phys. Rev. A} {\bf 76}, 042112 (2007).

\bibitem{finite-T3}
Obrecht, J. M.  \emph{ et al.}, Measurement of the Temperature Dependence of the Casimir-Polder Force. \emph{Phys. Rev. Lett.} {\bf 98}, 063201 (2007).

\bibitem{finite-T4}
Sushkov, A. O.,  Kim, W. J., Dalvit, D. A. R. \& Lamoreaux, S. K. Observation of the thermal Casimir force. \emph{Nature Physics} {\bf 7}, 230233 (2011). 










\bibitem{out-equilibrium1}
Antezza, M.,  Pitaevskii, L. P., Stringari, S. \& Svetovoy, V. B. Casimir-Lifshitz force out of thermal equilibrium. \emph{Phys. Rev. A} {\bf 77}, 022901 (2008).

\bibitem{out-equilibrium2}
Buhmann, S. Y. \& Scheel, S. Nonequilibrium thermal CasimirÐPolder forces. \emph{Phys. Scr.} {\bf T135} (2009) 014013.

\bibitem{out-equilibrium3}
Rodriguez, J. J. \& Salam, A. Casimir-Polder potential in a dielectric medium out of thermal equilibrium. \emph{Phys. Rev. A} {\bf 82}, 062522 (2010).

\bibitem{out-equilibrium4}
Noto, A., Messina, R., Guizal, B. \& Antezza, M. Casimir-Lifshitz force out of thermal equilibrium between dielectric gratings. \emph{Phys. Rev. A} {\bf 90}, 022120 (2014).

\bibitem{out-equilibrium5}
Zhou, W. \& Yu, H. Energy shift and Casimir-Polder force for an atom out of thermal equilibrium near a dielectric substrate. \emph{Phys. Rev. A} {\bf 90}, 032501 (2014).

\bibitem{out-equilibrium6}
Zhou, W. \& Yu, H. Casimir-Polder force for a polarizable molecule near a dielectric substrate out of thermal equilibrium. \emph{Phys. Rev. A} {\bf 91}, 052502 (2015).

\bibitem{out-equilibrium7}
Bimonte, G. Observing the Casimir-Lifshitz force out of thermal equilibrium. \emph{Phys. Rev. A} {\bf 92}, 032116 (2015).

\bibitem{Passante1}
Passante, R., Persico, F. \& Rizzuto, L. Causality, non-locality and three-body CasimirÐPolder energy between three ground-state
atoms. \emph{J. Phys. B: At. Mol. Opt. Phys.} {\bf 39} (2006) S685-694.

\bibitem{Passante2}
Rizzuto, L., Passante, R. \& Persico, F. Nonlocal Properties of Dynamical Three-Body Casimir-Polder Forces. \emph{Phys. Rev. Lett} {\bf 98}, 240404 (2007).

\bibitem{Cirone}
Cirone, M. A., Compagno, G., Palma, G. M., Passante, R. \& Persico, F. Casimir-Polder potentials as entanglement probe. \emph{EPL (Europhysics Letters)} {\bf 78} (2007) 30003.

\bibitem{Rizzuto1}
Rizzuto, L. Casimir-Polder interaction between an accelerated two-level system and an infinite plate. \emph{Phys. Rev. A} {\bf 76}, 062114 (2007).

\bibitem{Rizzuto2}
Marino, J., Noto, A. \& Passante, R. Thermal and Nonthermal Signatures of the Unruh Effect in Casimir-Polder Forces. \emph{Phys. Rev. Lett.} {\bf 113}, 020403, (2014).

\bibitem{Rizzuto3}
Rizzuto, L., \emph{et al.}, Non-thermal effects of acceleration in the resonance interaction between two uniformly accelerated atoms. arXiv:1601.04502 [quant-ph].

\bibitem{Zhang}
Zhang, J. \& Yu, H. Far-zone interatomic Casimir-Polder potential between two ground-state atoms outside a Schwarzschild black hole. \emph{Phys. Rev. A} {\bf 88}, 064501 (2013).

\bibitem{Tian}
Tian, Z. \& Jing, J. Distinguishing de Sitter universe from thermal Minkowski spacetime by Casimir-Polder-like force. \emph{JHEP} {\bf 07} (2014) 089.

\bibitem{Gibbons}
Gibbons, G. W. \& Hawking, S. W. Cosmological event horizons, thermodynamics, and particle creation. \emph{Phys. Rev. D} {\bf 15}, 2738 (1977).

\bibitem{Birrell}
Birrell, N. D. \& Davies, P. C. W. \textit{Quantum fields Theory in Curved Space}, (Cambridge University Press, Cambridge, England, 1982).

\bibitem{Polarski}
Polarski, D. On the Hawking effect in de Sitter space. \emph{Class. Quant. Grav.} {\bf 6} (1989) 717.



\bibitem{Deser}
Deser, S. \& Levin, O. Accelerated detectors and temperature in (anti-) de Sitter spaces. \emph{Class. Quant. Grav.} {\bf 14} (1997) L163.

\bibitem{Zhou-Yu1}
Zhu, Zh. \& Yu, H. Thermal nature of de Sitter spacetime and spontaneous excitation of atoms. \emph{JHEP} {\bf 02} (2008) 033.

\bibitem{Zhou-Yu2}
Zhou, W. \& Yu, H. Lamb shift in de Sitter spacetime. \emph{Phys. Rev. D} {\bf 82}, 124067 (2010).

\bibitem{Tian1}
Tian, Z. \& Jing, J. Geometric phase of two-level atoms and thermal nature of de Sitter spacetime. \emph{JHEP} {\bf 04} (2013) 109.

\bibitem{Tian2}
Tian, Z. \& Jing, J. Dynamics and quantum entanglement of two-level atoms in de Sitter spacetime. \emph{Ann. Phys.} {\bf 350} (2014) 1-13.

\bibitem{Entanglement-de-Sitter}
Steeg, G. V. \& Menicucci, N. C. Entangling power of an expanding universe. \emph{Phys. Rev. D} {\bf 79}, 044027 (2009).

\bibitem{Correlations-de-Sitter}
Nambu, Y. \& Ohsumi, Y. Classical and quantum correlations of scalar field in the inflationary universe. \emph{Phys. Rev. D} {\bf 84}, 044028 (2011).

\bibitem{Correlations-de-Sitter1}
Nambu, Y. Entanglement Structure in Expanding Universes. \emph{Entropy} {\bf 15}, 1847 (2013).

\bibitem{Other-means}
Hu, J. \& Yu, H. Quantum entanglement generation in de Sitter spacetime. \emph{Physical Review D} {\bf 88}, 104003 (2013).

\bibitem{Other-means1}
Salton, G., Mann, R. B. \& Menicucci, N. C. Acceleration-assisted entanglement harvesting and rangefinding. \emph{New J. Phys.} {\bf 17} 035001 (2015). 


\bibitem{resonance-books}
Craig, D. P. \& Thirunamachandran, T. \textit{Molecular Quantum Electrodynamics: An Introduction to Radiation-molecule Interactions}, (Dover Publ., Inc., Mineola, NY 1998).

\bibitem{resonance-books1}
Salam, A. \textit{Molecular Quantum Electrodynamics: Long-Range Intermolecular Interactions}, (John Wiley and Sons, Inc., Hoboken, New Jersey, 2010).

\bibitem{Benatti}
Benatti, F. \& Floreanini, R. Entanglement generation in uniformly accelerating atoms: Reexamination of the Unruh effect. \emph{Phys. Rev. A} {\bf 70}, 012112 (2004).


\bibitem{Benni}
R, Benni., Retzker, A. \& Silman, J. Violating BellÕs inequalities in vacuum. \emph{Phys. Rev. A} {\bf 71}, 042104 (2005). 

\bibitem{Yasusada}
Nambu, Y. \& Ohsumi, Y. Classical and quantum correlations of scalar field in the inflationary universe. \emph{Phys. Rev. D} {\bf 84}, 044028 (2011).

\bibitem{Yu1}
Yu, H. \& Zhang, J. Understanding Hawking radiation in the framework of open quantum systems. \emph{Phys. Rev. D} {\bf 77}, 024031 (2008).

\bibitem{Yu2}
Yu, H. Open Quantum System Approach to the Gibbons-Hawking Effect of de Sitter Space-Time. \emph{Phys. Rev. Lett} {\bf 106}, 061101 (2011).

\bibitem{Hu1}
Hu, J. \& Yu, H. Entanglement generation outside a Schwarzschild black hole and the Hawking effect. \emph{JHEP} {\bf 08}, (2011) 137.


\bibitem{Gorini}
Gorini, V., Kossakowski, A. \& Surdarshan, E. C. G. Completely positive dynamical semigroups of N-level systems. \emph{J. Math. Phys.} {\bf 17}, 821 (1976).

\bibitem{Lindblad}
Lindblad, G. On the generators of quantum dynamical semigroups. \emph{Commun. Math. Phys.} {\bf 48}, 119-130 (1976).

\bibitem{open}
Breuer, H.-P. \& Petruccione, F. \textit{The Theory of Open Quantum Systems}, (Oxford University Press, Oxford, 2002).

\bibitem{Dicke}
Dicke, R.H. Coherence in Spontaneous Radiation Processes. \emph{Phys. Rev.} {\bf 93}, 99 (1954).

\bibitem{Ficek}
Ficek, Z. \& Tana\'{s}, R. Entangled states and collective nonclassical effects in two-atom systems. \emph{Physics Reports} {\bf 372} (2002) 369-443.

\bibitem{Allen}
Allen, B. Vacuum states in de Sitter space. \emph{Phys. Rev. D} {\bf 32}, 3136 (1985).

\bibitem{Bethe1}
Bethe, H.A. The Electromagnetic shift of energy levels. \emph{Phys. Rev.} {\bf 72} (1947) 339.

\bibitem{Bethe2}
Welton, T.A. Some Observable Effects of the Quantum-Mechanical Fluctuations of the Electromagnetic Field. \emph{Phys. Rev.} {\bf 74} (1948) 1157.


\bibitem{Zhou1}
Zhou, W. \& Yu, H. The Lamb shift in de Sitter spacetime. \emph{Phys. Rev. D} {\bf 82} (2010) 124067; 

\bibitem{Zhou2}
Zhou, W. \& Yu, H. Lamb Shift for static atoms outside a Schwarzschild black hole. \emph{Phys. Rev. D} {\bf 82} (2010) 104030.


\bibitem{DDC}
Dalibard, J., Dupont-Roc, J. \& Cohen-Tannoudji, C. Vacuum fluctuations and radiation reaction: identification of tlieir respective contributions, \emph{J. Phys. (Paris)} {\bf 43} (1982) 1617.
\bibitem{DDC1}
Dalibard, J., Dupont-Roc, J. \& Cohen-Tannoudji, C. Dynamics of a small system coupled to a reservoir: reservoir fluctuations and self-reaction, \emph{J. Phys. (Paris)} {\bf 45} (1984) 637.


\end{thebibliography}
\end{document}